\documentstyle[aps,prl,graphicx]{revtex}

\input psfig.sty
\draft
\begin{document}
\twocolumn 
\wideabs{  
\title{Rapidly Rotating Bose-Einstein Condensates in and near the Lowest Landau Level}
\author{V. Schweikhard$^{1}$, I. Coddington$^{1}$, P. Engels$^{1}$, V. P. Mogendorff$^{2}$, and E.~A. Cornell$^{1}$\cite{qpdNIST}}
\address{$^{1}$ JILA, National Institute of Standards and Technology and University of Colorado,
and Department of Physics, University of Colorado, Boulder,
Colorado 80309-0440}
\address{$^{2}$ Physics Department, Eindhoven University of Technology, P.O. Box 513,\\
5600 MB Eindhoven, The Netherlands}
\date{\today}

\maketitle

\begin{abstract}
We create rapidly rotating Bose-Einstein condensates in the
lowest Landau level, by spinning up the condensates to rotation
rates $\Omega\,>\,99\%$ of the centrifugal limit for a
harmonically trapped gas, while reducing the number of atoms. As
a consequence, the chemical potential drops below the cyclotron
energy $2\hbar\Omega$. While in this mean-field quantum Hall
regime we still observe an ordered vortex lattice, its elastic
shear strength is strongly reduced, as evidenced by the observed
very low frequency of Tkachenko modes. Furthermore, the gas
approaches the quasi-two-dimensional limit. The associated
cross-over from interacting- to ideal-gas behavior along the
rotation axis results in a shift of the axial breathing mode
frequency.
\end{abstract}

} 

Rotating Bose-Einstein condensates (BECs) provide a conceptual
link between the physics of trapped gases and the physics of
condensed matter systems such as superfluids, type-II
superconductors and quantum Hall effect (QHE) materials. In all
these systems, striking counterintuitive effects emerge when an
external flux penetrates the sample. For charged particles this
flux can be provided by a magnetic field, leading to the
formation of Abrikosov flux line lattices in type-II
superconductors\cite{Abrikosov}, or in QHE systems to the
formation of correlated electron-liquids and composite
quasiparticles made of electrons with attached flux
quanta\cite{Laughlin}. For neutral superfluids, the analog to a
magnetic field is a rotation of the system, which similarly
spawns vortices\cite{Donelly}. In rotating atomic BECs, the
creation of large ordered Abrikosov lattices of vortices
\cite{Vortices} has recently become possible.
\par
Here we examine the vortex lattice of harmonically trapped BECs
approaching the high rotation limit, when the centrifugal force
quite nearly cancels the radial confining force. The formal
analogy of neutral atoms in this limit with electrons in a
strong magnetic field has led to the prediction that
quantum-Hall like properties should emerge in rapidly rotating
atomic BECs\cite{Gunn}. In particular, the single-particle
energy states organize into Landau levels, and if interactions
are weaker than the cyclotron energy, primarily the
near-degenerate states of the lowest Landau level (LLL) are
occupied. For rotating bosons in the LLL, three regimes have
been identified, distinguished by the filling factor $\nu\equiv
N_p/N_v$, i.e., the ratio of the number of particles ($N_p$) to
vortices ($N_v$). For high filling factors (always $\gtrsim 500$
in our system), the condensate is in the mean-field quantum Hall
regime\cite{Ho,Baym,BaymCores} and forms an ordered vortex
lattice ground state. With decreasing filling factor, the
elastic shear strength of the vortex lattice decreases, which is
reflected in very low frequencies of long-wavelength transverse
lattice excitations (Tkachenko
oscillations\cite{Anglin,JILATkachenko,BaymTkachenko,MacDonaldPC,TkachenkoSimulations}).
For filling factors below $\nu \approx 10$ the shear strength is
predicted to drop sufficiently for quantum fluctuations to melt
the vortex lattice\cite{Gunn,MacDonald}, and a variety of
strongly correlated vortex liquid states similar to those in the
Fermionic fractional QHE are predicted to appear\cite{Gunn}. For
still lower $\nu$ exotic quasiparticle excitations obeying
fractional statistics\cite{Cirac} are predicted.
\par
In this Letter we report the observation of rapidly rotating
BECs in the lowest Landau level, and provide evidence that the
elastic shear strength of the vortex lattice drops substantially
as the BEC enters the mean-field quantum Hall
regime\cite{BaymTkachenko}. This effect is a precursor to the
predicted quantum melting of the lattice at lower filling
factors. Our rapidly rotating condensates spin out into a
pancake shape and approach the quasi-two-dimensional regime. We
observe a corresponding cross-over in the spectrum of breathing
excitations along the axial direction. The high rotation limit
has been studied experimentally in
Ref.\cite{DalibardCriticalRot}, focusing on effects of a
rotating trap anisotropy, which is not present in our setup, and
in Ref.\cite{ENSFastRot} where the addition of a quartic term to
the trapping potential led to a loss of vortex visibility.
\par
Our experiments take place in an axially symmetric harmonic trap
with oscillation frequencies
$\{\omega_{\rho},\omega_{z}\}=2\pi\{8.3,5.3\}\,\rm{Hz}$. Using
an evaporative spin-up technique described in Ref.
\cite{GiantVortex} we create condensates
containing up to $5.5\times 10^6$  $^{87}Rb$ atoms in the
$|F=1,m_{F}=-1\rangle$ state, rotating about the vertical,
z~axis, at a rate $\tilde{\Omega} = 0.95$ ($\tilde{\Omega}
\equiv\Omega/\omega_{\rho}$ is the rotation rate $\Omega$ scaled
by the centrifugal rotation limit $\omega_{\rho}$ for a
harmonically trapped gas). To further approach the limit
$\tilde{\Omega}\rightarrow 1$, we employ an optical spin-up
technique, where the BEC is illuminated uniformly with laser
light, and the recoil from spontaneously scattered photons
removes atoms from the condensate. Since the condensate is
optically thin to the laser light, atoms are removed without
position or angular-momentum selectivity, such that angular
momentum per particle is unchanged. Atom loss leads to a small
decrease in cloud radius which, through conservation of angular
momentum, increases $\tilde{\Omega}$. Over a period of up to 2
seconds we decrease the number of BEC atoms by up to a factor of
$100$, to $5\times10^4$, while increasing $\tilde{\Omega}$ from
$0.95$ to more than $0.99$\cite{RotationFootnote}. At this
point, further reduction in number degrades the quality of
images unacceptably. Ongoing evaporation is imposed to retain a
quasi-pure BEC with no discernible thermal cloud.
\par
\begin{figure}
\begin{center}
\psfig{figure=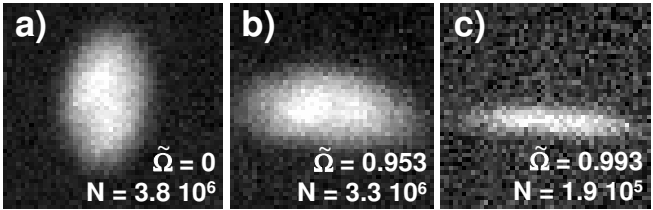,width=1\linewidth,clip=}
\end{center}
\caption {Side view images of BECs in trap. (a)
Static BEC. The aspect ratio $R_z/R_{\rho}=1.57$ ($N = 3.8\times
10^6$ atoms) resembles the prolate trap shape. (b) After
evaporative spin-up, $N = 3.3\times 10^6$, $\tilde{\Omega} =
0.953$, (c) evaporative plus optical spin-up, $N=1.9\times
10^5$, $\tilde{\Omega} = 0.993$. Due to centrifugal distortion
the aspect ratio is changed by a factor 8 compared to (a).}
\label{SpinupSideview}
\end{figure}

With increasing rotation, centrifugal force distorts the cloud
into an extremely oblate shape [see Fig.\ref{SpinupSideview}]
and reduces the density significantly - thus the BEC approaches
the quasi-two-dimensional regime. For the highest rotation rates
we achieve, the chemical potential $\mu$ is reduced close to the
axial oscillator energy,
$\Gamma_{2D}\equiv\frac{\mu}{2\hbar\omega_z}\approx 1.5$, and
the gas undergoes a cross-over from interacting- to ideal-gas
behavior along the axial direction. To probe this cross-over, we
excite the lowest order axial breathing mode over a range of
rotation rates. For a BEC in the axial Thomas-Fermi regime, an
axial breathing frequency $\omega_B=\sqrt{3}\,\omega_z$ has been
predicted in the limit $\tilde{\Omega}\rightarrow
1$\cite{Stringari}, whereas $\omega_B =2\,\omega_z$ is expected
for a non-interacting gas.
\par
\begin{figure}
\begin{center}
\psfig{figure=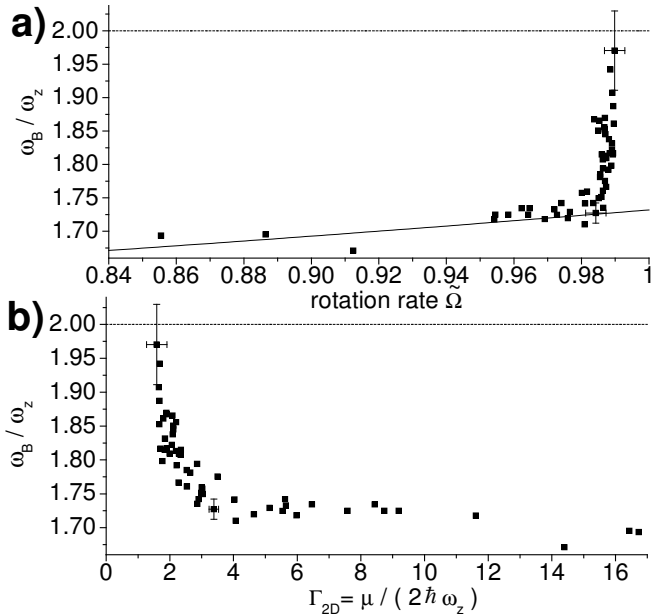,width=1\linewidth,clip=}
\end{center}
\caption {Measured axial breathing frequency
$\omega_B/\omega_z$ (a) as a function of rotation rate
$\tilde{\Omega}$ and (b) vs. $\Gamma_{2D}$. Solid line:
Prediction for the hydrodynamic regime [20]; Dashed line: ideal
gas limit. For $\tilde{\Omega}>0.98$ ($\Gamma_{2D}<3$) a
cross-over from interacting- to ideal-gas behavior is observed.
Representative error bars are shown for two data points.}
\label{Axialbreathing}
\end{figure}

To excite the breathing mode, we jump the axial trap frequency
by $6\%$, while leaving the radial frequency unchanged (within
$< 0.5\%$). To extract the axial breathing frequency $\omega_B$,
we take 13 nondestructive in-trap images of the cloud,
perpendicular to the axis of rotation. From the oscillation of
the axial Thomas-Fermi radius\cite{TFGaussFootnote} in time we
obtain $\omega_B$. Rotation rates are
obtained\cite{RotationFootnote} from the aspect ratio by
averaging over all 13 images to eliminate the effect of axial
breathing. As shown in Fig. \ref{Axialbreathing}(a), we do
indeed observe a frequency cross-over from
$\omega_B=\sqrt{3}\,\omega_z$ to $\omega_B = 2\,\omega_z$  as
$\tilde{\Omega}\rightarrow 1$. To quantify under which
conditions the cross-over occurs, we plot the same data vs.
$\Gamma_{2D}$ [Fig.
\ref{Axialbreathing}(b)], where the chemical potential is
determined from the measured atom number, the rotation rate and
the trap frequencies. For $\Gamma_{2D}< 3$, the ratio
$\omega_B/\omega_z$ starts to deviate from the predicted
hydrodynamic value, and approaches $2$ for our lowest
$\Gamma_{2D}\approx 1.5$.
\par
As $\tilde{\Omega} \rightarrow 1$, also the dynamics in the
radial plane are affected. For the highest rotation rates,
interactions become sufficiently weak that the chemical
potential $\mu$ drops below the cyclotron energy $2\hbar\Omega$,
which is only a few percent smaller than the Landau level
spacing $2\hbar\omega_{\rho}$. Then,
$\Gamma_{LLL}\equiv\frac{\mu}{2
\hbar\Omega}<1$, and the condensate primarily occupies
single-particle states in the LLL. These form a ladder
of near-degenerate states, with a frequency splitting of
$\epsilon=\omega_{\rho}-\Omega$. The number of occupied states
is $N_{LLL}\approx \frac{\mu}{\hbar
\epsilon}$. We are able to create condensates with $\Gamma_{LLL}$ as
low as $0.6$, which occupy $N_{LLL}\approx 120$ states with a
splitting $\epsilon < 2 \pi \, 0.06\,\rm{Hz}$. In this regime of
near-degenerate single-particle states a drastic decrease of the
lattice's elastic shear strength takes place. The elastic shear
modulus $C_2$ is predicted by Baym\cite{BaymTkachenko} to
decrease with increasing rotation rate from its value in the
``stiff" Thomas-Fermi (TF) limit, $C_2^{TF}=n_{(\Omega)}
\hbar\Omega/8$ (where $n_{(\Omega)}$ is the BEC number density)
to its value in the mean-field quantum Hall regime, of
$C_2^{LLL}\approx 0.16\times\Gamma_{LLL}\times C_2^{TF}$. We
directly probe this shear strength by exciting the lowest order
azimuthally symmetric lattice mode ( $(n=1,m=0)$ Tkachenko
mode\cite{Anglin,JILATkachenko,BaymTkachenko}). Its frequency
$\omega_{(1,0)} \sim
\sqrt{C_2}$ is expected to drop by a factor $\approx 2.5$ below the
TF prediction when $\Gamma_{LLL} = 1$.
\par
\begin{figure}
\begin{center}
\psfig{figure=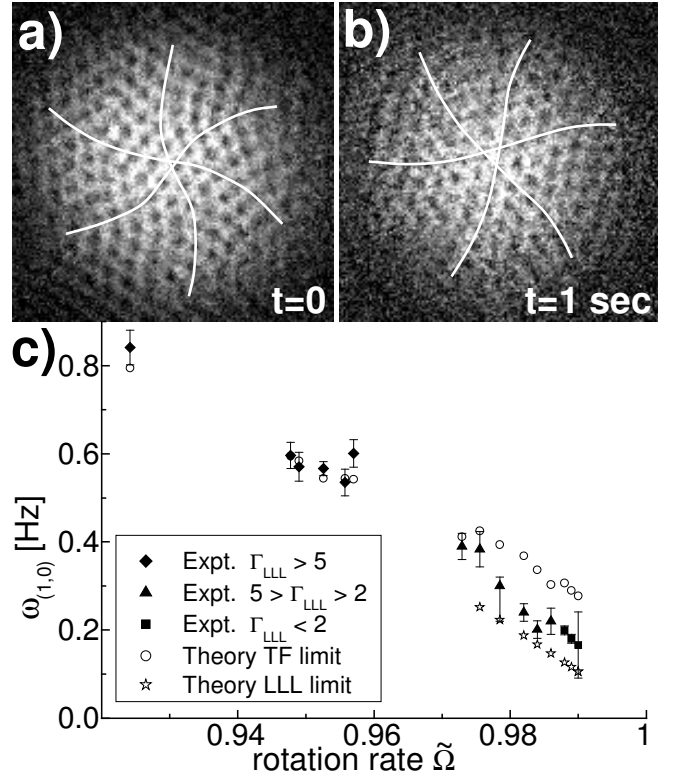,width=1\linewidth,clip=}
\end{center}
\caption{ (a), (b) Tkachenko mode at
$\Gamma_{LLL} = 1.2$ ($N=1.5\times 10^5$, $\tilde{\Omega} =
0.989$): (a) directly after excitation, (b) after 1 sec - the
lattice oscillation has not yet completed $1/4$ cycle. (c)
Comparison of measured Tkachenko mode frequency $\omega_{(1,0)}$
(solid symbols) vs. $\tilde{\Omega}$ to theory [11], using
vortex lattice shear modulus $C_2^{TF}$ in the Thomas-Fermi (TF)
limit (circles), and $C_2^{LLL}$ in the mean-field quantum Hall
regime (stars). Note that both N and $\Gamma_{LLL}$ decrease as
$\tilde{\Omega}$ increases. For $\Gamma_{LLL}\approx 3$ (reached
at $N=7.8\times 10^5$, $\tilde{\Omega}\approx 0.978$) the data
cross over from the TF to the quantum Hall prediction.}
\label{Tkachenko}
\end{figure}

Our excitation technique for Tkachenko modes has been described
in Ref. \cite{JILATkachenko}. In brief, we shine a focused red
detuned laser ($\lambda = 850
\,\rm{nm}$) onto the BEC center, along the axis of
rotation. This laser draws atoms into the center, and Coriolis
force diverts the atoms' inward motion into the lattice rotation
direction. The vortex lattice adjusts to this distortion, and
after we turn off the beam, the lattice elasticity drives
oscillations at the frequency $\omega_{(1,0)}$. We observe the
oscillation by varying the wait time after the excitation, and
then expanding the condensate before imaging the vortex lattice
along the z-axis [see Fig. \ref{Tkachenko}(a),(b)]. In Fig.
\ref{Tkachenko}(c), we compare the measured frequencies
$\omega_{(1,0)}$ to the predictions of Ref.\cite{BaymTkachenko}
for the TF limit and for the mean-field quantum Hall regime. For
$\tilde{\Omega} < 0.98$ ($\Gamma_{LLL} > 3$), the frequency
$\omega_{(1,0)}$ follows the prediction for the TF regime,
whereas by $\tilde{\Omega} = 0.990$ ($\Gamma_{LLL} = 1.5$),
$\omega_{(1,0)}$ has dropped to close to the prediction for the
LLL, thus providing evidence for the cross-over to the lower
shear modulus $C_2$ predicted for the LLL.
\par
While we are able to produce clouds with $\Gamma_{LLL}$ measured
to be substantially lower than $\Gamma_{LLL} = 1.5$, we are
unable to accurately measure Tkachenko frequencies under these
extreme conditions, due at least in part to the very weakness of
$C_2$. The Tkachenko mode frequencies become so low that it
takes multiple seconds to track even a quarter oscillation [Fig.
\ref{Tkachenko}(a),(b)], while at the same time the very weak shear strength means
that even minor perturbations to the cloud can cause the lattice
to melt and the individual cores to lose
contrast\cite{LossDalibard} in a matter of seconds. These
perturbations can result from residual asymmetry of the magnetic
trapping potential, or from spatial structure in the optical
beam used to reduce the atom number, or perhaps from thermal
fluctuations. In contrast, for a ``stiff" cloud of $3\times
10^6$ atoms at $\tilde{\Omega} = 0.95$ ($\Gamma_{LLL}\approx7$)
we observe that the lattice remains ordered, and
$\tilde{\Omega}$ can be kept constant, over the entire $1/e$
lifetime of the BEC ($\approx 3$ minutes).
\par
Finally we discuss observations related to two other predicted
consequences of rapid rotation. First, the radial condensate
density profile has been predicted to change from the parabolic
Thomas-Fermi (TF) profile to a Gaussian profile in the
LLL\cite{Ho}. We create clouds with varying values of
$\Gamma_{LLL}$ down to as small as $0.6$, and after $3\,\rm sec$
equilibration time acquire images of the radial density profile.
These images fit better to a TF profile than to a Gaussian,
showing no signs of a cross-over in the radial density profile
as the LLL is entered. On the theoretical side, it has been
pointed out\cite{MacDonaldPC} that, whereas the predictions of
Ref.\cite{Ho} are based on an assumption that the areal density
of vortices is uniform, a very small deviation from uniformity
could account for an overall TF density profile. Furthermore,
according to Ref.\cite{BaymCores}, a Gaussian radial profile is
to be expected only if $Na/R_z \ll 1$ ($N$: number of atoms in
the BEC, $a$: scattering length, $R_z$: axial BEC radius),
whereas our measurements were performed on condensates with
$Na/R_z\gtrsim 26$. It certainly seems reasonable to expect a TF
profile for a given direction when the chemical potential is
much greater than the confining frequency in that direction, and
for our condensates with $\Gamma_{LLL}=0.6$, the chemical
potential $\mu/\hbar\approx2\pi\,10\,\rm{Hz}$ while the
(centrifugally weakened) effective radial trap frequency
$\omega_{eff}=\sqrt{\omega_{\rho}^2-\Omega^2}\approx
2\pi\,1\,\rm{Hz}$.
\par
Second, there has been a debate about the possibility of vortex
core overlap in the high rotation
limit\cite{Baym,BaymCores,Feder}. To achieve adequate imaging
resolution, we expand the BEC before imaging the vortex lattice.
Our expansion procedure\cite{GiantVortex} leads to nearly pure
two-dimensional (2-d) expansion (a thirteen-fold expansion in
radius occurs while the axial size changes by $\lesssim 25\%$).
We expect that under these conditions, the ratio of the vortex
core area to the lattice's unit cell area should be preserved
throughout the expansion process. We fit a 2-d Gaussian to the
missing optical density associated with each vortex core. The
vortex radius $r_v$ is defined to be the RMS radius of the 2-d
Gaussian, which is $0.60$ times its FWHM. We then define the
fractional area $\mathcal{A}$ occupied by the vortices to be
$\mathcal{A}$ $=n_v
\pi r_v^2$, where $n_v$ is the areal density of vortices. In the
limit of many vortices, the theoretical prediction for $n_v$ is
$m\Omega/(\pi \hbar)$.
\par
\begin{figure}
\begin{center}
\psfig{figure=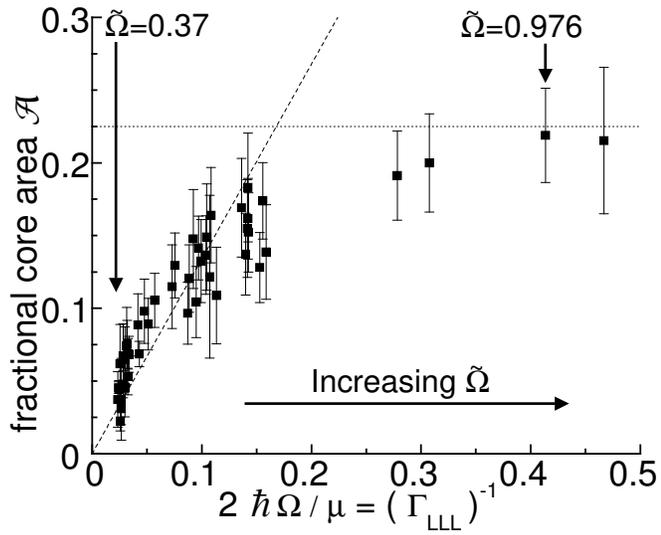,width=1\linewidth,clip=}
\end{center}
\caption {Fraction of the condensate surface area occupied by
vortex cores, $\mathcal{A}$, measured after condensate expansion
(squares), plotted vs. the inverse of the lowest Landau level
parameter, $(\Gamma_{LLL})^{-1}=2\hbar\Omega/\mu$. The data
clearly show a saturation of $\mathcal{A}$, as
$\tilde{\Omega}\rightarrow 1$. Dashed line: prediction (see
text) for the pre-expansion value at low rotation rate. Dotted
line: result of Ref.[8] for the saturated value of $\mathcal{A}$
in the LLL.}
\label{VortexCores}
\end{figure}

To determine a theoretical value for the vortex core size, we
perform a numerical simulation of the Gross-Pitaevskii equation
of a BEC containing an isolated vortex. We obtain $r_v =
1.94\times \xi$ with $\xi=(8\pi n a)^{-\frac{1}{2}}$, where $a$
is the scattering length and $n$ is the
density\cite{Densityfootnote}. The resulting prediction for
$\mathcal{A}$ can be expressed as $\mathcal{A}$ $=
1.34\times(\Gamma_{LLL})^{-1}$. This value exceeds unity for
$\Gamma_{LLL}< 1.34$, which has led to the prediction that
vortices should merge in the LLL limit. An alternate treatment
due to Baym and coworkers\cite{BaymCores} on the other hand
predicts that $\mathcal{A}$ saturates at $0.225$ in the LLL. Our
data for $\mathcal{A}$ are plotted in Fig. \ref{VortexCores}.
For $\Gamma_{LLL}^{-1}<0.1$ the data agree reasonably well with
our numerical result. For larger $\Gamma_{LLL}^{-1}$ the data
clearly show saturation of $\mathcal{A}$ at a value close to the
LLL limit\cite{BaymCores}, rather than a divergence of
$\mathcal{A}$ as $\tilde{\Omega}\rightarrow 1$. Further details
on vortex core structure will be provided in a future
publication\cite{Vortexcores}.
\par
In conclusion, we have created rapidly rotating BECs in the
lowest Landau level. The vortex lattice remains ordered, but its
elastic shear strength is drastically reduced. In
expansion-images we find no divergence of vortex core area as
$\tilde{\Omega}\rightarrow 1$, as well as no deviation from a
radial Thomas-Fermi profile. Additionally, our rapidly rotating
BECs approach the quasi-two-dimensional limit. We remain far
from the regime in which quantum fluctuations\cite{MacDonald}
should destroy the lattice, but observing the effects of thermal
fluctuations\cite{Shlyapnikov,BaymCorrelationfct} in this
reduced-dimensionality system may be possible.
\par
We would like to acknowledge G. Baym, C. Pethick, J. Sinova, J.
Diaz-Velez, C. Hanna, A. MacDonald, N. Read and D. Feder for
useful discussions and calculations. This work was funded by NSF
and NIST. V.M. acknowledges financial support from the
Netherlands Foundation for Fundamental Research on Matter (FOM).


%

\begin{thebibliography}{21}
\bibitem{RotationFootnote}
Rotation rates are accurately determined by comparing the
measured BEC aspect ratio to the trap aspect ratio (see
e.g.\cite{Stringari}). At our lowest values of $\Gamma_{2D}$ we
correct for quantum-pressure contributions to axial size.
\bibitem{TFGaussFootnote}
For the smallest values $\Gamma_{2D}$, the axial density profile
is slightly better fitted by a Gaussian. To avoid bias however
we fit all data assuming a TF profile.
\bibitem{LossDalibard}
Melting and loss of contrast are also reported in Ref.$[17]$.
\bibitem{Densityfootnote}
The density $n=7/10 \,n_{peak}$ is density-weighted along the
rotation axis and is radially averaged over the vortex cores
within $1/2$ the TF radius, as only in this region we fit the
observed vortex cores.
\end{thebibliography}
\end{document}